# G-corrected holographic dark energy model

**M. Malekjani[1,2] • M. Honari-Jafarpour[1]**

**Abstract**

Here we investigate the holographic dark energy model in the framework of FRW cosmology where the newtonian gravitational constant,$G$, is varying with cosmic time. Using the complementary astronomical data which support the time dependency of $G$, the evolutionary treatment of EoS parameter and energy density of dark energy model are calculated in the presence of time variation of $G$. It has been shown that in this case, the phantom regime can be achieved at the present time. We also calculate the evolution of $G$-corrected deceleration parameter for holographic dark energy model and show that the dependency of $G$ on the comic time can influence on the transition epoch from decelerated expansion to the accelerated phase. Finally we perform the statefinder analysis for $G$- corrected holographic model and show that this model has a shorter distance from the observational point in $s-r$ plane compare with original holographic dark energy model.

## 1 Introduction

Nowadays, it is widely believed that the cosmos is experiencing an accelerated expansion. This idea and belief came into existence after collection of data from "Type Ia supernova" in 1998 (Perlmutter et al 1998). Also the other data from WMAP (Bennett et al 2009),SDSS (Tegmark et al 2004) and X-ray (Allen et al 2004) experiments support this accelerated expansion. In the framework of standard cosmology, the existence of dark energy with negative pressures is essential to interpret the cosmic acceleration. Hence, dark energy scenario has got a lot of attention in modern cosmology both from theoretical and observational point of view. Observationally, The result of SNeIa experiment shows that dark energy occupies about 72% of the total energy of our universe, dark matter and baryons about 28% of the total energy of the universe(Perlmutter et al 1998). Although the nature of dark energy is still un-known, but the ultimate fate of the current universe is determined by this mysterious component. Till now, some theoretical models have been proposed to interpret the behavior of the dark energy.The first and the simplest one is Einstein's "cosmological constant" (Sahni & Starobinsky 2003) which, of course, has two problems called fine -tuning and cosmic coincidence. The cosmological constant has the fixed equation of state $w_\Lambda = -1$, while the dynamics of current expansion can be explained by dynamical dark energy models with time varying equation of state. The scalar fields such as quintessence (Wetterich 1998), phantom (Caldwell 2002) or the combination of both which is called quintom (Elizalde et al 2004) are examples of dynamical models. The other dynamical dark energy models which interprets the current acceleration of expansion are constructed based on quantum gravity theory(Witten 2002). The holographic dark energy (HDE) model is one of the the proposed models based on the holographic principle in quantum gravity (Horava & Minic 2000). According to the holographic principle, a short distance ultra-violet (UV) cut-off is related to the long distance infra-red (IR)

M. Malekjani
M. Honari-Jafarpour
E-mail: malekjani@basu.ac.ir.
E-mail: M.Honarijafarpour@basu.ac.ir.
[1]Department of Physics, Faculty of Science, Bu-Ali Sina University, Hamedan 65178, Iran.
[2]Research Institute for Astronomy & Astrophysics of Maragha (RIAAM)- Maragha, Iran, P. O. Box: 55134-441.

cut-off, due to the limit set by the formation of a black hole (Horava & Minic 2000). The holographic principle indicates that the zero-point energy of a system with size L should not be exceed from the mass of black hole with the same size. From the above principle, the energy density of HDE model in cosmology can be described as:

$$\rho_d = \frac{3c^2}{8\pi G L^2} \tag{1}$$

Where L is the cosmic horizon and c is a numerical constant of order unity and G is a Newton's gravitational constant. The length scale $L$ has an essential role in the definition of energy density of HDE model. Therefore the various model of HDE have been constructed for different of infrared (IR) cutoff length. For example the simple choice of IR cutoff is the Hubble length which leads to wrong equation of state for DE (Horava & Minic 2000). However in the presence of interaction between dark matter and DE the HDE model with Hubble radios IR cutoff can derive the accelerated expansion and also solve the coincidence problem (Pavon & Zimdahl 2005). The other choice for IR cutoff is the particle horizon. In this case, like Hubble length, the accelerated expansion cannot be achieved Pavon & Zimdahl (2005). Another choice is the event horizon where the cosmic acceleration can be interpreted in this case (Zhou et al 2007). Nojiri and Odintsov (2006) investigated the holographic DE model by assuming IR cutoff depends on the Hubble rate, particle and future horizons, span of life of the universe and cosmological. In this generalized form of HDE the phantom regime can be achieved and also the coincidence problem is demonstrated. Unification of early phantom inflation and late time acceleration of the universe is the other feature of this model. Recently, the HDE model has been constrained by various astronomical observations(Huang & Gong 2004; Zhang & Wu 2005; Wu et al. 2008; Enqvist et al. 2005).

In addition, there are some theoretical and observational supports indicating that Newton's gravity constant varies and changes with cosmic time. The first theoretical idea in this respect is the pioneering work of Dirac(Dirac 1938), and then the idea of Dyson(Dyson 1972). Also, the Branse-Dicke framework in the Physics predicts the variability and fluctuation of G(Brans & Dicke 1961). Moreover the varying behavior of $G$ in Kaluza-Klein theory was associated with a scalar field appearing in the metric component corresponding to the 5-th dimension and its size variation(Kaluza, et al 1921).In this theory, a scalar field paired with gravity by a new parameter replaces the quantum gravity. The variability of $G$ with time is also supported from the observational viewpoint. The observational data collected by Type Ia Supernova (Gaztanaga et al 2002),Hulse-Taylor Binary(Damour et al 1988), astro-seismological data from pulsating white dwarf stars (Benvenuto 2004; Biesiada & Malec 2004), helio-sesmiological (Guenther 1998) and the Big Bang Nuclei-synthesis data(Copi et al 2004) support a variable value for G with time. We refer to these observations in the section 3 of the paper. Here in this work we consider the HDE model with time varying $G$, the so-called $G$-corrected HDE model, in spatially flat FRW universe. We consider the event horizon as an IR cut-off in relation (1). In this concern some other works have been done in which the HDE model has been considered with time dependency of $G$, i.e., (Jiano et al 2009). Here we obtain the equation of state $w_d$ as well as deceleration parameter $q$ and statefinder pair $\{s, r\}$ for $G$-corrected HDE model in FRW universe and also solve the related equations numerically by using the observational values for $G(t)$.

It is clear that constraining a given model against the observational data is model dependent. Therefore some doubts usually remain on the validity of the constraints on cosmological parameters. In order to solve this problem, we use the cosmography, i.e. the expansion of scale factor in Taylor series with respect to the cosmic time. The first term of Taylor series is the Hubble parameter ($H = \frac{da}{adt}$), the second term is the deceleration parameter ($q = -\frac{d^2a}{aH^2dt^2}$), the third term is the jerk parameter ($r = \frac{d^3a}{aH^3dt^3}$), the forth term is snap ($k = \frac{d^4a}{aH^4dt^4}$) and the fifth term is lerk parameter ($l = \frac{d^5a}{aH^5dt^5}$). The present values of the above parameters can be used to describe the evolution of the universe. For example $q_0 < 0$ indicates the current accelerated expansion of the universe and also $r_0$ allows to discriminate between different dark energy models. Since Hubble's parameter which corresponds to the first derivative of the scale factor ($\dot{a}$) and the deceleration parameter which corresponds to the second derivative of the scale factor ($\ddot{a}$) can not distinguish between the different models, we need a higher derivative of scale factor. Sahni et al.(Sahni et al 2003)and Alam et al.(Alam et al 2003b), by using the third time derivative of scale factor, introduced the statefinder pair {s, r} in order to diagnosis the treatment of dark energy models. The statefinder pair in spatially flat universe is given by:

$$r = \frac{\dddot{a}}{aH^3} \quad ; \quad s = \frac{r-1}{3(q-\frac{1}{2})} \tag{2}$$

The statefinder parameters s and r are the geometrical parameters, because they only depend on the scale factor. Up to now, different dark energy models have been investigated from the viewpoint of statefinder diagnostic. These models have different evolutionary trajectories in {s, r} plane, therefore the statefinder tool can discriminate these models. The well known $\Lambda - CDM$ model corresponds to the fixed point $\{s = 0, r = 1\}$ in the $s - r$ plane (Sahni et al 2003). The distance of the current value of statefinder pair {s,r} of a given dark energy model from the fixed point $\{s = 0, r = 1\}$ is a valuable criterion to examine of model. Also the recant investigation by observational data resulted the best fit value for statefinder in flat universe as $\{s_{obs} = -0.006, r_{obs} = 1.02\}$ (M. Malekjani & Khodam-Mohammadi 2012c). The other dark energy models which have been studied from the viewpoint of statefinder diagnostic are :
the quintessence DE model (Sahni et al 2003; Alam et al 2003b) , the interacting quintessence models (Zimdahl & Pavon 2004; Zhang 2005a), the holographic dark energy models (Zhang 2005b; Zhang et al 2007) , the holographic dark energy model in non-flat universe (Setare et al 2007), the phantom model (Chang et al 2007), the tachyon (Shao & Gui 2007), the generalized chaplygin gas model (Malekjani et al 2011a), the interacting new agegraphic DE model in flat and non-flat universe (Zhang 2010; Khodam-Mohammadi & Malekjani 2010), the agegraphic dark energy model with and without interaction in flat and non-flat universe (Wei & Cai 2007; Malekjani & Khodam-Mohammadi 2010), the new holographic dark energy model (Malekjani et al 2011b), the interacting polytropic gas model (Malekjani & Khodam-Mohammadi 2012a) and the interacting ghost dark energy model (M. Malekjani & Khodam-Mohammadi 2012b).
The paper is organized as follows: In section (2) the $G$-corrected HDE model has been presented in falt FRW universe and the equation of sate $w_d$, deceleration parameter $q$ and statefinder pair $\{s, r\}$ have been calculated in the presence of time variation of $G$. In section (3) we present the numerical results and in section (4) the paper is concluded.

## 2 The G-corrected HDE model in a FRW cosmology

The Hilbert-Einstein action with time varying gravitational constant, $G(t) = G_0\phi(t)$, is

$$S = \frac{1}{16\pi G_0}\int\sqrt{-g}[\frac{R}{\phi(t)} + L_m]d^4x \quad (3)$$

Here we assume the scalar function $\phi(t)$ for time dependency of $G(t)$. Also $G_0$ is usual gravitational constant and $L_m$ is the lagrangian of matter field. By variation of above action with respect to metric $g_{\mu\nu}$ the first corrected Friedmann equation for zero-zero component of field equation in flat geometry can be obtained as follows

$$H^2 = \frac{8\pi G(t)}{3}(\rho_m + \rho_d) + H\frac{\dot{G}}{G} \quad (4)$$

Since the value of $\dot{G}/G$ is small particularly in the late time accelerated universe, therefore we have ignored the higher time derivative of $G$ (i.e., $\ddot{G}/G$) and also larger powers than one (i.e., $(\dot{G}/G)^2$, ...).
The equation (4) for standard model with time varying gravitational constant can also be obtained from Branse- Dicke gravity by assuming ($w = 0$ and $\psi = 1/\phi(t)$) in equation (2) of (Banerjee & Pavon 2007). Here $w$ is the Branse-Dicke parameter and $\psi$ is Branse-Dicke scalar field.
If we consider the derivative of G according to $\ln a$ the above $G$-corrected Friedman equation can be re-written as:

$$H^2(1 - \frac{\acute{G}}{G}) = \frac{8\pi G(t)}{3}(\rho_m + \rho_d), \quad (5)$$

where prime is derivative with respect to $x = \ln a$.

Assuming the event horizon as an IR cut-off as

$$R_h = a\int\frac{dt}{a} = a\int\frac{H}{\acute{a}}d\acute{a}, \quad (6)$$

The energy density of HDE model in Eq.(1) is written as

$$\rho_d = \frac{3c^2}{8\pi G(t)R_h^2} \quad (7)$$

In terms of dimensionless energy densities

$$\Omega_m = \frac{\rho_m}{\rho_c} \quad ; \quad \Omega_d = \frac{\rho_d}{\rho_c}, \quad (8)$$

where the$\rho_c = \frac{3H^2}{8\pi G(t)}$ is the critical energy density, the corrected Friedman equation(5) can be written as

$$\Omega_m + \Omega_d = 1 - \frac{\acute{G}}{G} \quad (9)$$

this equation is look like to the Friedman equation in the non-flat universe : $\Omega_m + \Omega_d = 1 - \Omega_k$. Based on observational data described in introduction we consider the negative values for $\frac{\acute{G}}{G}$. Therefore the added term $\acute{G}/G$ in right hand side of (9) can be interpreted as non-flatness parameter $\Omega_k$ in non-flat universe.
In addition the evolution of Hubble parameter in terms of scale factor in $G$-corrected flat universe including dark matter and dark energy can be calculated from Eq.(4) as follows

$$H^2(1-\frac{\acute{G}}{G}) = {H_0}^2[\Omega_m a^{-3} + \Omega_d a^{-3(1+w_d)}], \quad (10)$$

where $H_0$ is the present value of Hubble parameter.

The conservation equations for dark matter and dark energy are given by:

$$\dot{\rho}_m + 3H\rho_m = 0 \quad (11)$$

$$\dot{\rho}_d + 3H(1+w_d)\rho_d = 0 \quad (12)$$

Taking the time derivative of (7) by using $\dot{R}_h = 1+HR_h$ and (7) in relation (12) we obtain the equation of state for G-corrected HDE model as follows

$$w_d = -\frac{1}{3} - \frac{2}{3}\frac{\sqrt{\Omega_d}}{c} + \frac{1}{3}\frac{\acute{G}}{G} \quad (13)$$

Also, taking the derivative of (13) with respect to $x = \ln a$, we obtain

$$\acute{w_d}(1-\frac{1}{2}\frac{\acute{G}}{G}) = \frac{-1}{3}\times \Big(\frac{\acute{\Omega_d}}{c\sqrt{\Omega_d}}(1-\frac{\acute{G}}{2G}) - \frac{3}{2}(1+w_d\Omega_d)\frac{\acute{G}}{G}\Big) \quad (14)$$

Here we have ignored the terms including $(\acute{G}/G)^2$ and $(\acute{G}/G)^3$ and also $\acute{\acute{G}}/G$. In what follows which we derive and calculate, we keep only the first- order correction of $G$ (i.e., $\acute{G}/G$).

Now, derivative of $\Omega_d = \frac{\rho_d}{\rho_c} = \frac{c^2}{H^2R_h^2}$ yields the evolutionary equation for dark energy density for G-corrected HDE model as follows

$$\acute{\Omega_d} = -2\Omega_d[\frac{c}{HR} + \frac{\dot{H}}{H^2} + 1] \quad (15)$$

In addition taking the time derivative of corrected Friedman equation (4) obtains

$$\frac{\dot{H}}{H^2}(1-\frac{1}{2}\frac{\acute{G}}{G}) = -\frac{3}{2}(1+w_d\Omega_d) + 2\frac{\acute{G}}{G} \quad (16)$$

Therefore the equation of motion for energy density of $G$-corrected HDE, i.e., Eq.(15) is written as

$$\acute{\Omega_d} \quad (1-\frac{\acute{G}}{2G}) = \Omega_d \times \Big(3(1+w_d\Omega_d) + \frac{\sqrt{\Omega_d}}{c}(2-\frac{\acute{G}}{G}) - 3\frac{\acute{G}}{G} - 2\Big) \quad (17)$$

The deceleration parameter $q = -1 - \dot{H}/H^2$ which represents the decelerated or accelerated phase of the expansion of the universe, by using (13)and (16), is written for G-corrected HDE model as

$$q(1-\frac{1}{2}\frac{\acute{G}}{G}) = \frac{1}{2}(1+3w_d\Omega_d) - \frac{3}{2}\frac{\acute{G}}{G} \quad (18)$$

For completeness, we now derive the statefinder pair $\{s,r\}$ in G-corrected HDE model. For this aim, by time derivative of (16), we first obtain

$$\frac{\ddot{H}}{H^3} \quad (1-\frac{3}{2}\frac{\acute{G}}{G}) = \frac{9}{2}(1+w_d\Omega_d)\Big(w_d\Omega_d(1-\frac{3}{4}\frac{\acute{G}}{G}) - \frac{11}{4}\frac{\acute{G}}{G} + 1\Big) - \frac{3}{2}(1-\frac{\acute{G}}{G})(\acute{w_d}\Omega_d + \acute{\Omega_d}w_d) \quad (19)$$

Inserting (16) and (19) in $r = \frac{\ddot{H}}{H^3} + 3\frac{\dot{H}}{H^2} + 1$ we obtain the following equation for the parameter $r$ of statefinder pair

$$r \quad (1-\frac{3}{2}\frac{\acute{G}}{G}) = \frac{9}{2}(1+w_d\Omega_d)\Big(w_d\Omega_d(1-\frac{3}{4}\frac{\acute{G}}{G}) - \frac{7}{4}\frac{\acute{G}}{G}\Big) - \frac{3}{2}(1-\frac{\acute{G}}{G})(\acute{w_d}\Omega_d + \acute{\Omega_d}w_d) + \frac{9}{2}\frac{\acute{G}}{G} + 1 \quad (20)$$

From (2), by using (18) and (20) we also obtained the parameter $s$ in G-corrected HDE model as follows

$$s = \Big[\frac{3}{2}(1+w_d\Omega_d)\Big(w_d\Omega_d(1-\frac{5}{4}\frac{\acute{G}}{G}) - \frac{7}{4}\frac{\acute{G}}{G}\Big) - \frac{1}{2}(1-\frac{3}{2}\frac{\acute{G}}{G})(\acute{w_d}\Omega_d + \acute{\Omega_d}w_d) + 2\frac{\acute{G}}{G}\Big] / \Big[\frac{3}{2}w_d\Omega_d(1-\frac{3}{2}\frac{\acute{G}}{G}) - \frac{5}{4}\frac{\acute{G}}{G}\Big] \quad (21)$$

In the limiting case of time-independent gravitational constant G (i.e., $\acute{G} = 0$) all the above relations reduce to those obtained for original holographic dark energy (OHDE) model in (Zhang 2005).

## 3 Numerical result

There are many astronomical observations which show the time dependency of Newtonian gravitational constant. All these data are in agreement with Dyson idea who pointed out that $G$ varies in the length of cosmic age $H^{-1}$. Based on the observational data from WMAP five-year observations the present value of Hubble parameter is $H_0 = 6.64\times10^{-11}yr^{-1}$ (Bennett et al 2009; Zhang & Wu 2009). Moreover the astronomical observations are in the line of Dirac's theory in which $G(t) \propto t^{-1}$ as follows (Cetto et al 1986)

$$G(t) = k_1 H(t) = k_2[H(t)]^{\frac{2}{3}}\rho(t)^{-\frac{1}{2}} \quad (22)$$

where $k_1$ and $k_2$ are constant. The data gathered from SNeIA data yields the best rang for variation of $G$ as: $-10^{-11}yr^{-1} \leq \frac{\dot{G}}{G} \leq 0$ (Gaztanaga et al 2002) and the data obtained from Binary Pulsar PSR1913 determines the range of variation of $\frac{\dot{G}}{G}$ as: $-(1.10 \pm 1.07) \times 10^{-11}yr^{-1} < \frac{\dot{G}}{G} < 0$ (Damour et al 1988). The data obtained from Helio-sesmiological determines the best range $-1.6 \times 10^{-12}yr^{-1} < \frac{\dot{G}}{G} < 0$ (Guenther 1998).

Another estimation for $\frac{\dot{G}}{G}$ has been done through astro-seismological data obtained from pulsating white dwarf star which yields the best range of variation as: $-2.5\times10^{-10}yr^{-1} \leq \frac{\dot{G}}{G} \leq +4.5\times10^{-10}yr^{-1}$ (Benvenuto 2004). In (Biesiada & Malec 2004), the range of $\frac{\dot{G}}{G}$ was determined as $\frac{\dot{G}}{G} \leq +4.1 \times 10^{-11}yr^{-1}$. It should be noted that all the above range of $\frac{\dot{G}}{G}$ are calculated for $z \leq 3.5$. Finally from the observational data of Big Bang nuclei-synthesis, we have $-4.0\times10^{-13}yr^{-1} < \frac{\dot{G}}{G} < +3 \times 10^{-13}yr^{-1}$ (Copi et al 2004). More details for the variation of $G$ with cosmic time can be seen in (Ray & Mukhopadhyay 2007). In previous section we calculated the effect of variation of $G$ on the HDE model in terms of $\frac{G'}{G}$. Therefore, we change the time derivative to derivative with respect to $x = \ln a$ as $\frac{\dot{G}}{G} = H\frac{G'}{G}$ where $\frac{G'}{G}$ is a dimensionless numerical value, because the dimensions of Hubble Parameter is same as $\frac{\dot{G}}{G}$. Here we call this numerical value as $\alpha = \frac{G'}{G}$. In this work we use the SNeIa observational data$-10^{-11}yr^{-1} \leq \frac{\dot{G}}{G} \leq 0$ which covers the other observational range of $\frac{\dot{G}}{G}$ . We also use the present value $H_0 = 6.64\times10^{-11}yr^{-1}$ based on observational data from WMAP five-year observations (Bennett et al 2009; Zhang & Wu 2009). The parameter $\alpha$, using by these observational data can be obtained as $|\alpha| \sim 0.10$. Therefore we choose the illustrative values $\alpha = -0.1, 0, 0.1$ which are in the order of the observational value. At follows we calculate the evolution of cosmological quantities: EoS parameter, energy density, deceleration parameter and statfinder pair of G-corrected HDE model and obtain the effect of parameter $\alpha$ on the evolution of these cosmological quantities.

### 3.1 EoS parameter

By solving (13), we show the evolution of EoS parameter of G-corrected HDE as a function of redshift in Fig.(1). Here we fix the holographic parameter $c = 0.87$. Note that for this value the original HDE model without $G$ correction can not enter the phantom regime. The black solid curve relates to original HDE model without $G$ correction. The red- dashed curve is indicated for $\alpha = 0.1$ and blue- dotted- dashed line represents $\alpha = 0.1$. Here we see that the G-corrected HDE model can enter to phantom regime when $\alpha < 0$, i.e. blue-dashed line. Hence one can conclude that the G-corrected HDE model can cross the phantom divide without a need of interaction between dark matter and dark energy. Also, the G-corrected HDE model crosses that phantom line ($w_d = -1$) from up ($w_d > -1$) to below ($w_d < -1$). This behavior of G-corrected HDE model is in agreement with recent observations in which the universe transits from quintessence regime ($w_d > -1$) to the phantom regime ($w_d < -1$) at the near past (Alam et al 2004).

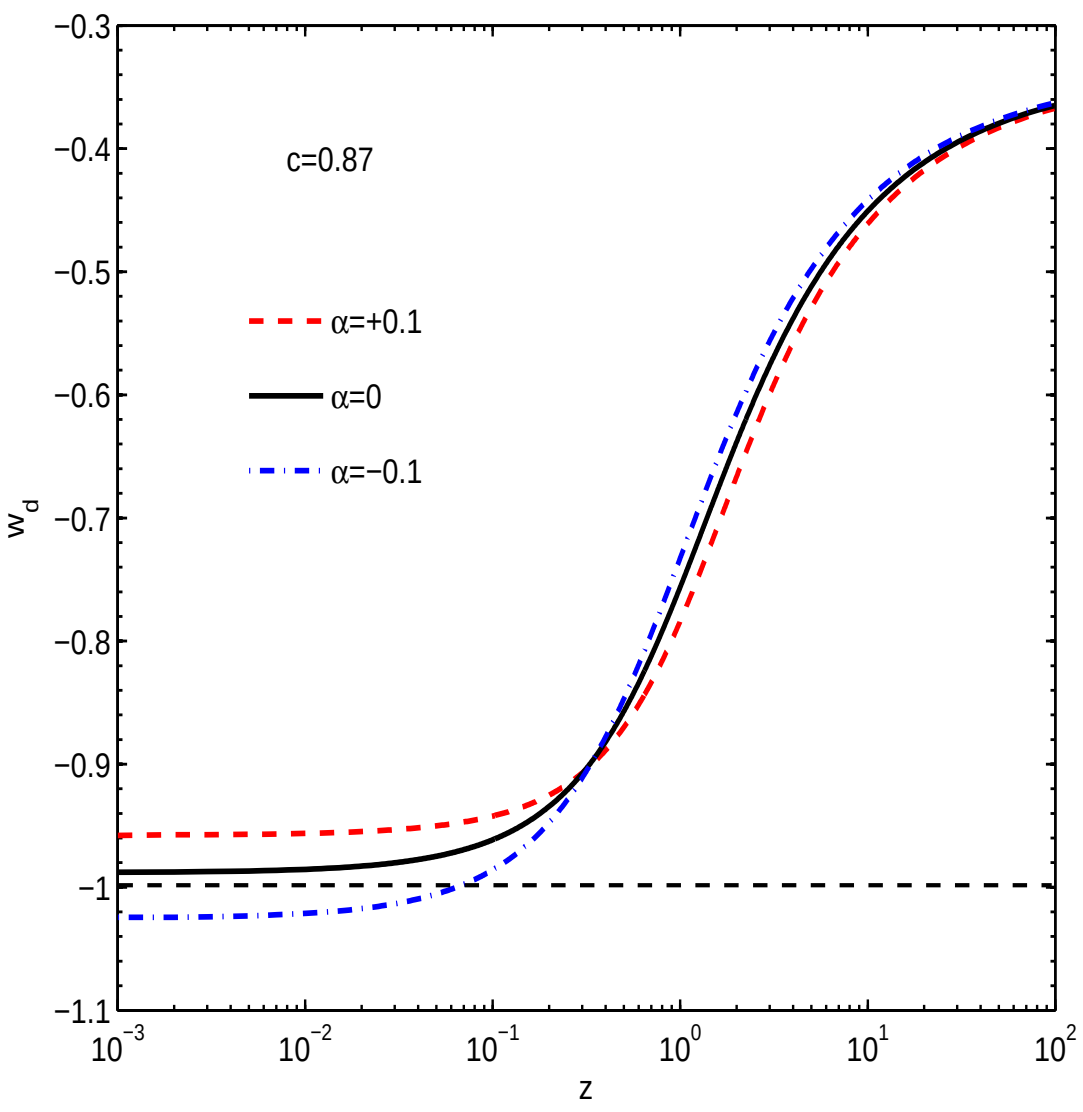

**Fig. 1** The evolution of EoS parameter of G-corrected HDE model versus redshift parameter $z$ for different illustrative values of $\alpha$ as indicated in legend.

### 3.2 energy density

Here we calculate the evolution of energy density of G-corrected HDE model as a function of redshift parameter from the early time up to late time by solving equation (15). In Fig.(2), we plot the evolution of energy density $\Omega_d$ versus of redshift for different illustrative values of $\alpha$. We see that at the early times $\Omega_d \to 0$ and at the late times $\Omega_d \to 1$, meaning the dark energy dominated universe at the late time. In this figure by fixing $c = 0.87$ the parameter $\alpha$ is varied as illustrative values $-0.1, 0.0, +0.1$ corresponding to dotted-dashed -blue, solid -black and dashed -red curves , respectively. We see that in the past times the dark energy becomes more dominant for positive values of $\alpha$ and at the late times the dark energy dominated universe can be achieved sooner for negative values.

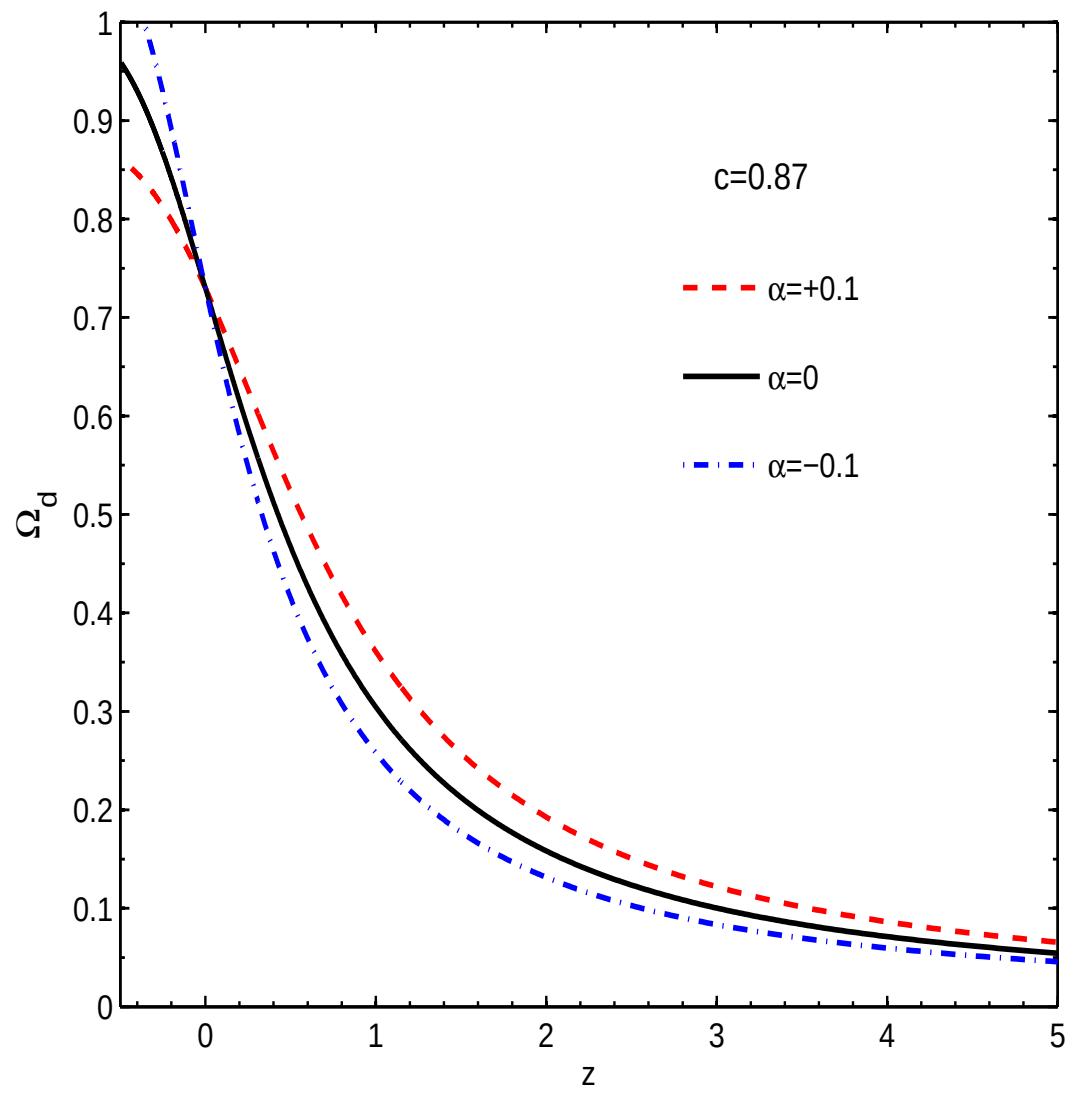

**Fig. 2** The evolution of density parameter of Dark energy of G-corrected HDE model($\Omega_d$) versus redshift parameter $z$ for different illustrative values of $\alpha$. We can see the different value of $\alpha$ result the different evolutionary trajectory in terms of redshift

### 3.3 deceleration parameter

Here we study the expansion phase of the universe by calculating the evolution of deceleration parameter $q$ in G-corrected HDE model. By solving equation (18) and using (15), we plot the evolution of $q$ versus redshift parameter $z$ in Fig.(3). We see that the parameter $q$ start from $q = 0.50$, representing the $CDM$ model at the early time. Then the parameter $q$ becomes negative, representing the accelerated expansion phase of the universe at recent epochs. Therefore the G-corrected HDE model can interpret the decelerated phase of the expansion of the universe at the early times and accelerated phase later. we fix the parameter $c = 0.8$ and for the different illustrative value of the $\alpha = -0.1, 0.0, +0.1$ corresponding to dotted-dashed -blue, solid -black and dashed -red curves , respectively. We see for negative value of $\alpha$, the accelerated expansion can be achieved sooner than the original HDE model($\alpha = 0.0$) and also positive value of $\alpha$.

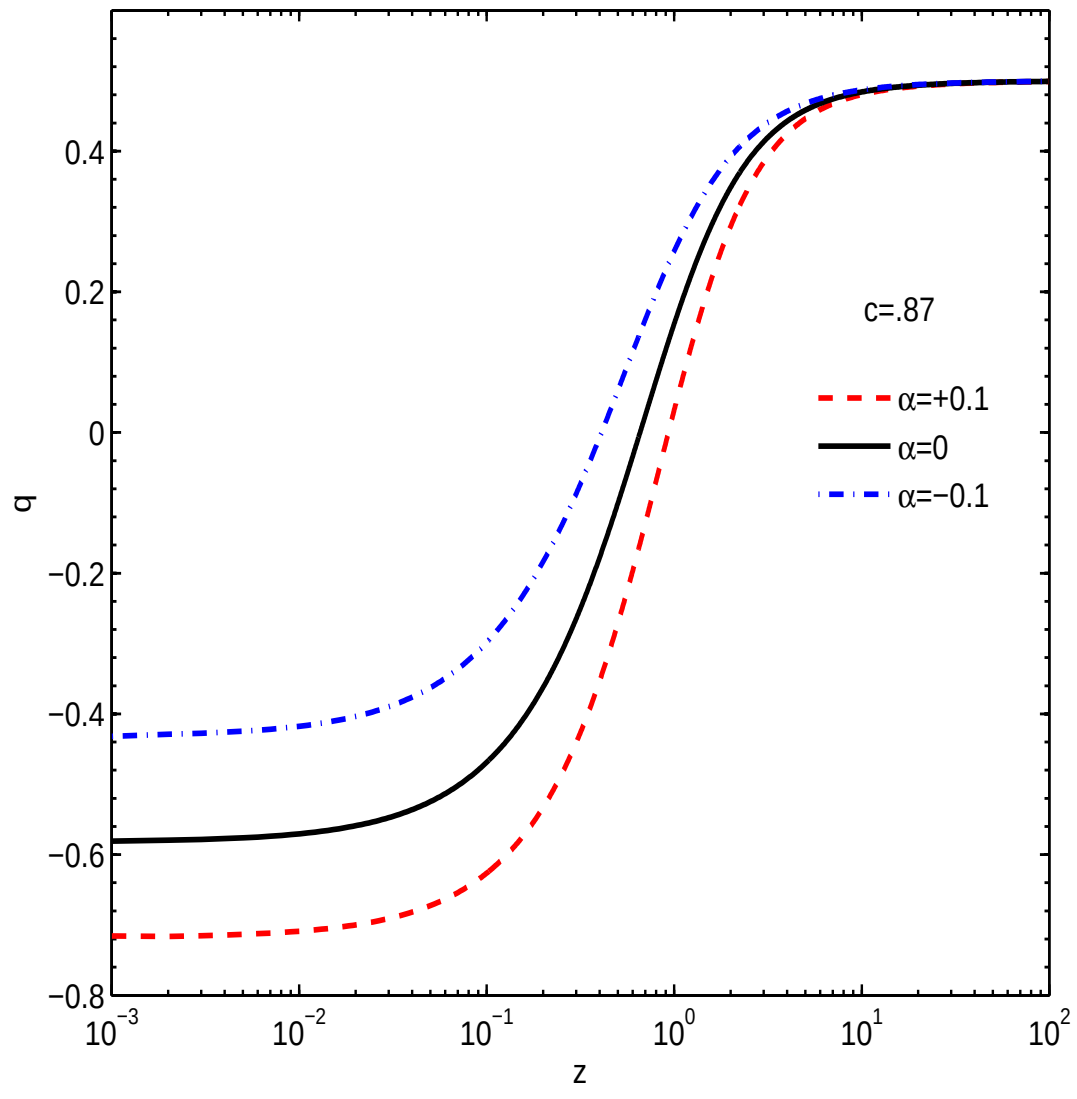

**Fig. 3** The evolution of deceleration parameter ($q$) of G-corrected HDE model as a function of redshift parameter $z$ for different illustrative values of $\alpha$. We can see accelerated expansion can be achieved sooner for $\alpha = +0.1$

### 3.4 statefinder diagnosis

The statefinder pair {s, r} for G-corrected HDE model is given by relations (20) and (21). In statefinder plane, the horizontal axis is defined by the parameter $s$ and vertical axis by the parameter $r$. In Fig.(4), by putting (13),(14)and (15)in (20) and (21) and solving them ,we obtain the evolutionary trajectories of G-corrected HDE model in $s-r$ plane for different values of parameter $\alpha$. By expanding the universe, the evolutionary trajectories evolve from right to left. The parameter $r$ decreases then increases, while the parameter $s$ decreases forever. The trajectories cross the $\Lambda-CDM$ fixed point $\{s=0, r=1\}$ at the near past. In the other words, the G-corrected HDE model has mimicking the $\Lambda$CDM model at the near past. The present values of the cosmographic parameters, introduced in introduction, have been observationally constrained using the Markov Chain Monte Carlo method in (Capozziello et al 2011) as follows: $H_0 = 0.718$, $q_0 = -0.64$, $r_0 = 1.02$ , $k_0 =?0.39$, $l = 4.05$ . Using $q_0 = -0.64$ and $r_0 = 1.02$, we calculate the present value of statefinder parameter $s$ as $s_0 = -0.006$. Hence the observational point $s_0 = -0.006, r_0 = 1.02$ in s-r diagram is very close to $\Lambda$CDM fixed point $s_0 = 0, r_0 = 1$. The observational point is indicated by green star in Fig. (4). Here we fix the holographic parameter $c = 0.87$ and vary $\alpha$ as $\alpha = -0.1, 0.0, +0.1$ corresponding to dotted-dashed -blue, solid -black and dashed -red curves, respectively. We see that different values of $\alpha$ result different trajectories in $s-r$ plane. Therefore the G-corrected HDE model in $s-r$ plane is discriminated for different values of $\alpha$. The colored circles on the curves represent the today's value of statefinder parameters $\{s_0, r_0\}$ of the model. We also see that for positive values of $\alpha$, the distance of $\{s_0, r_0\}$ from the observational point $\{s_{obs} = -0.006, r_{obs} = 1.02\}$ is shorter and for negative values of $\alpha$ and longer for positive values of $\alpha$ compare with original HDE model.

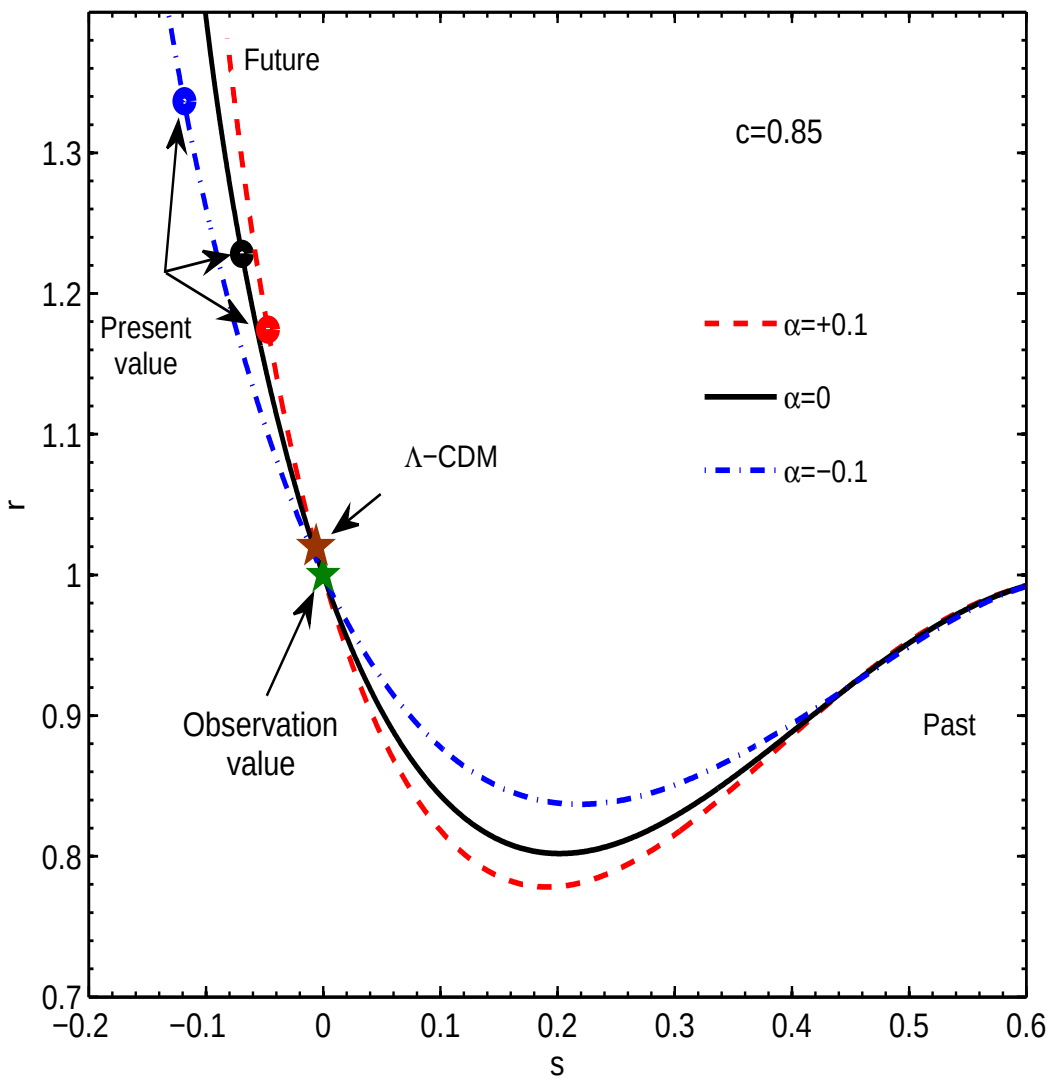

**Fig. 4** The he evolutionary trajectories of G-corrected HDE model in $s-r$ plane for different values of $\alpha$.we can see the different value of $\alpha$ result different evolutionary trajectories. also we can see for the $\alpha = +0.1$ the distance of present value from $\Lambda - CDM$ fixed point and $\{s_{obs}, r_{obs}\}$(present value) are shorter

## 4 conclusion

In summary, we extended the holographic dark energy (HDE) model by assuming the time dependency of Newtonian gravitational constant, $G$, in standard model of cosmology. Here we obtained the $G$-corrected Friedman equation in flat universe. Regarding, the astronomical data from type Ia Supernova (Gaztanaga et al 2002),Hulse-Taylor Binary (Damour et al 1988), astro-seismological data from pulsating white dwarf stars (Benvenuto 2004; Biesiada & Malec 2004), helio-seismological data (Guenther 1998) and the Big Bang Nuclei-synthesis data(Copi et al 2004), we obtained the parameter $|\alpha| = \frac{G'}{G} = 0.10$. The evolution of EoS parameter, deceleration parameter and energy density parameter of HDE model in the presence of $G$ correction have been calculated. We showed that the $G$ correction can affect the evolution of above parameters at the present time up to near past and is negligible at the early times. It was shown that for an illustrative value of holographic parameter $c$ in which the original HDE model can not cross the phantom line, the $G$- corrected HDE model can achieve the phantom regime and cross the phantom line from up ($w_d > -1$) to below ($w_< -1$) in agreement with recent observations (Alam et al 2004). The parameter $\alpha$ can also influence on the transition from decelerated expansion to the accelerated expansion. We showed that for $\alpha > 0$ the transition from $q > 0$ to $q < 0$ earlier and for $\alpha < 0$ later compare with original HDE model. Finally we performed the statefinder diagnosis analysis for $G$-corrected HDE model and showed that the $G$ correction can affect on the evolutionary trajectories of the model in $s-r$ plane. We concluded that for $\alpha > 0$, the distance of present value $\{s_0, r_0\}$ from the observational point is shorter and for $\alpha < 0$ is longer compare with original HDE model.

**Acknowledgment**
We are grateful to A. Khodam-Mohammadi for helpful discussions.